\documentclass{jfm}
\usepackage{fix-cm}
\usepackage{graphicx}
\usepackage{epstopdf,epsfig}
\usepackage{tikz,amssymb}
\usepackage{newtxtext}
\usepackage{newtxmath}
\usepackage{natbib}
\usepackage{hyperref}
\usepackage{placeins}
\usepackage[normalem]{ulem}
\hypersetup{
    colorlinks = true,
    urlcolor   = blue,
    citecolor  = black,
}

\newcommand{\RomanNumeralCaps}[1]
\linenumbers

\newcommand{\Rt}{\delta^+}

\definecolor{r1}{HTML}{6495ED} 
\definecolor{r2}{HTML}{F08080} 
\definecolor{r3}{rgb}{0.8,0.4,0.0} 
\definecolor{ra}{HTML}{00008B} 

\newcommand{\goodymarker}{
  \tikz[baseline=-0.2ex]\draw[fill=red,draw=black]
  (0,0)--(1ex,0)--(.5ex,1ex)--cycle;}

\title{{Two-component inner--outer scaling} model for the wall-pressure spectrum at high Reynolds number}

\author{J. M. O. Massey \aff{1}
  \corresp{\email{masseyj@stanford.edu}},
  A. J. Smits\aff{2}, \and B. J. McKeon\aff{1}}

\affiliation{\aff{1}Center for Turbulence Research, Stanford University, Stanford, CA 94305, USA
  \aff{2}Department of Mechanical and Aerospace Engineering, Princeton University, Princeton, NJ 08544, USA}

\begin{document}
\maketitle

\begin{abstract}
    Wall-pressure fluctuations beneath turbulent boundary layers drive noise and structural fatigue through interactions between fluid and structural modes.  Conventional predictive models for the spectrum--such as the widely accepted Goody model (\textit{AIAA Journal} 42 (9), 2004, 1788--1794)--fail to capture the energetic growth in the {low-frequency range} that occurs at high Reynolds number, while at the same time over-predicting the variance.
    To address these shortcomings, two semi-empirical models are proposed for the wall-pressure spectrum in canonical turbulent boundary layers, pipes and channels for friction Reynolds numbers $\delta^+$ ranging from 180 to 47 000.
    Consistent with the approach outlined modelling the streamwise Reynolds stress in the recent work of Gustenyov et al. (\textit{J. Fluid Mech.} 1016, 2025, A23), the models are based on consideration of two {spectral components} that represent the contributions to the wall pressure fluctuations from inner-scale motions and outer-scale motions.
    The first model expresses the pre-multiplied spectrum as the sum of two overlapping log-normal {components}: an inner-scaled term that is $\delta^+$-invariant and an outer-scaled term whose amplitude broadens smoothly with $\delta^+$. Calibrated against large-eddy simulations, direct numerical simulations,  and recent high-$\delta^+$ pipe data, it reproduces the {inner-scaled peak} and the emergence of an outer-scaled peak at large $\delta^+$.
    The second model, developed around newly available pipe data, uses theoretical arguments to prescribe the spectral shapes of the inner and outer {components}. Embedding the $\delta^+$-dependence in smooth asymptotic functions yields a formulation that varies continuously with $\delta^+$ {and generalises beyond the calibration range}. Both models capture the full spectrum and {recover} the observed logarithmic growth of its variance, {providing a compact, physics-informed empirical representation} for more accurate engineering predictions of wall-pressure fluctuations.
\end{abstract}


\section{Introduction}
    Predicting radiated noise and mitigating structural resonance in aircraft and marine structures depend critically on accurate models of wall-pressure behaviour in turbulent wall-bounded flows. A complete description of the fluctuating wall-pressure field is given by the three-dimensional wavenumber-frequency spectrum, $\phi_{pp}(f,k_x,k_z)$ \citep{zhao_review_2024}. In canonical incompressible wall-bounded flows--e.g. zero-pressure gradient (ZPG) boundary layers and smooth-wall internal flows--this spectrum exhibits a convective ridge and a low-frequency range. The convective ridge maps closely to Taylor's frozen turbulence hypothesis \citep{taylor1938spectrum}, but shows weak scale-dependence \citep{del_alamo_estimation_2009}, with a noticeable reduction from the typical inner-normalised mean convection velocity $U_c^+\equiv U_c/u_\tau \approx 10$ at high wavenumbers, where $u_\tau$ is the friction velocity.

    {In most experiments, however, only the one-dimensional frequency spectrum $\phi_{pp}(f)$ is available. Because $\phi_{pp}(f)$ is obtained by integrating over wavenumbers (and implicitly over a range of scale-dependent convection velocities), features in $\phi_{pp}(f)$ cannot be uniquely attributed to specific regions of $k$--$\omega$ space. In this paper we therefore focus on the scaling behaviour and evolution of the 1-D spectral features.}

    {This restriction--although full-aperture experimental arrays are becoming more ubiquitous (e.g. \citet{damani_measurement_2025})--has motivated a family of semi-empirical models that reconstruct $\phi_{pp}(f,k_x,k_z)$ from $\phi_{pp}(f)$ \citep{corcos_structure_1964,smol2006new,hwang_comparison_2009}.} Consequently, the fidelity of $\phi_{pp}(f)$ as a function of the friction Reynolds number, $\Rt$, directly governs the accuracy of predicted wall-pressure behaviour and underpins efforts to scale its variance \citep{farabee_spectral_1991,hu_wall_2006-1,klewicki_statistical_2008,schlatter_assessment_2010,lee_direct_2015,panton_correlation_2017,hasan2025scaling}.

    A widely used model for the wall-pressure spectrum in zero-pressure-gradient boundary layers is that proposed by \cite{goody_empirical_2004}. {Other models are available, but are limited by their use of low-mid $\Rt$ datasets; see \citet{damani_measurement_2025} for a recent comparison of available models}. The Goody model is derived from $\{\mathit{Re}_{\theta_i}\}_{i=1}^{7}\subset [1.4\times10^{3},\,2.34\times10^{4}] \mapsto \{\Rt_i\}_{i=1}^{7}\subset[650,\,7\,650]$.  It encapsulates distinct inner- and outer-time-scales and echoes Bradshaw's early recognition of inner- and outer-scaled contributions \citep{bradshaw_inactive_1967}.  Its key assumption is an overlap region in which dimensional analysis predicts an $f^{-1}$ scaling. Recent high-$\Rt$ measurements reveal, however, an outer-scaled spectral peak that violates this simple $f^{-1}$ behaviour and leads to errors in the Goody model at high-Reynolds number \citep{klewicki_statistical_2008,fritsch_pressure_2020,fritsch_fluctuating_2022,gibeau_low-_2021,damani_evaluating_2024,Damani_Butt_Totten_Devenport_Lowe_2025,dacome_scaling_2025}.

\begin{figure}
        \centering
        \includegraphics{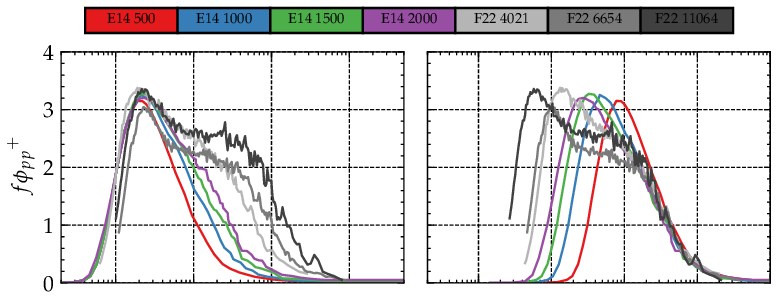} \\ \vspace{-1mm}
        \includegraphics{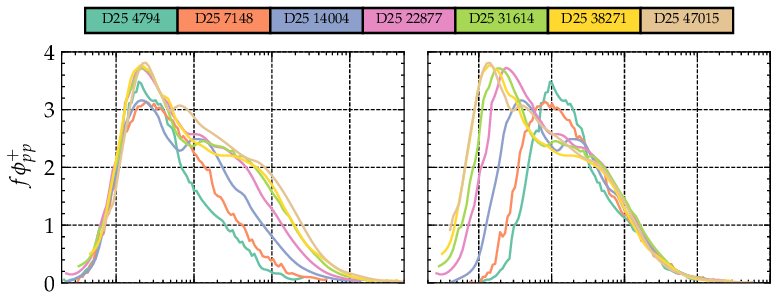} \\ \vspace{-1mm}
        \includegraphics{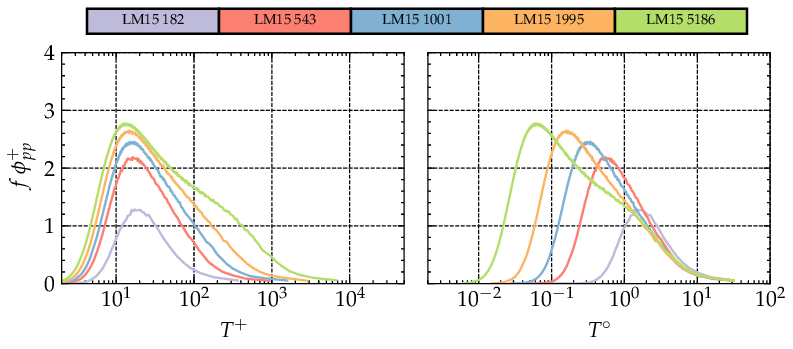}\\ \vspace{-3.5mm}
        \caption{Pre-multiplied spectra of wall-pressure fluctuations.  Left column:  inner scaling.  Right column: outer scaling.   Top row:  Boundary layers.  Highly-resolved LES data from \cite{eitel-amor_simulation_2014} for $\Rt=500$ to 2000,  experimental  data from \cite{fritsch_pressure_2020,fritsch_fluctuating_2022} for $\Rt=4021$  to 11,064.  Middle row: pipes. Experiments by  \cite{dacome_scaling_2025} for $\Rt=4794$  to 47,015.  Bottom row:  channels. DNS data from \cite{lee_direct_2015} for $\Rt=180$ to 5200.}
        \label{fig:data}
    \end{figure}

    To develop models that correctly capture both the high Reynolds number behaviour of the spectrum, and the Reynolds-number dependence of the variance, we use data from boundary-layer, pipe and channel flows over a very wide range of Reynolds numbers $\{\Rt_i\}_{i=1}^{19}\subset[180,\,4.7\times10^4]$ (figure~\ref{fig:data}).  In particular, we exploit the diagnostic power of the pre-multiplied spectrum, $f\phi_{pp}$, which more clearly separates inner- and outer-scale contributions than the conventional log-log representation. The inner-scaled spectrum is $\phi_{pp}^{+}= \phi_{pp}/\tau_{w}^{2}$, so the pre-multiplied form is ${f\phi_{pp}}^{+}$ and the variance is $\langle p_{w}^{2}\rangle^{+}=\int_{0}^{\infty}{f\phi_{pp}}^{+}\,\mathrm{d}\log f$.  Superscript $(\cdot)^{+}$ denotes normalisation by the viscous length $\nu/u_\tau$, while superscript $(\cdot)^{o}$ denotes normalisation by $\delta$, the 99\% boundary layer thickness or pipe/channel half-height, and $U_e$, the freestream {or centreline} velocity . Frequency and period are related by $f=1/T$, while $T^{+}=Tu_\tau^{2}/\nu$ and $T^{o}=TU_{e}/\delta$ are the inner- and outer-scaled periods, respectively.  The boundary-layer data are taken from highly-resolved large-eddy simulations (LES) \citep{eitel-amor_simulation_2014} and experiments \citep{fritsch_pressure_2020,fritsch_fluctuating_2022}; the pipe flow data are from the CICLoPE facility \citep{dacome_scaling_2025}; and the channel flow data are from DNS \citep{lee_direct_2015}. Further details on the data are given in Appendix A. 
    
    We see that the pre-multiplied spectra all share similar features. First, the data collapse at low values of $T^+$ (high frequencies) in inner scaling, {consistent with the findings of \citet{pirozzoli_pressure_2025}, who note the universality at the small-scale end of the spectra}. In outer scaling at high values of $T^o$ (low frequencies), the spectra also collapse.   Second, there is a peak located at $T^+ \approx 10-15$ for channels and $T^+ \approx 20$ for boundary layers and pipes. This peak is identified with the start of the {inner-scaled peak} and its magnitude varies with Reynolds number at low Reynolds numbers, more so for the internal flows than for the boundary layer.   Third, as the Reynolds number increases, there is increased energy content at low frequencies ($T^o = O(1)$), marking the development of the {low-frequency portion}.

    In what follows, we use these observations to develop two semi-empirical models for the wall-pressure spectrum that explicitly represent the bimodal picture as a sum of inner- and outer-scaled {spectral components}. The models apply to boundary layer, pipe flow, and channel flows, and {capture the pre-multiplied spectra} at high Reynolds number while reproducing the Reynolds-number-dependent behaviour of the variance in agreement with previous work.

\section{Modelling Approach}\label{sec:model approach}
    \begin{figure}
        \centering
        \includegraphics{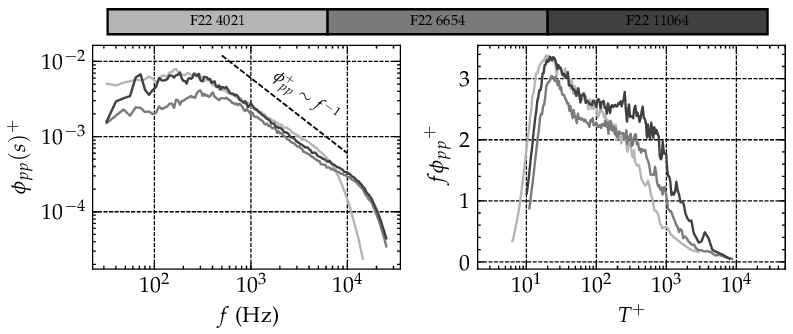}
        \caption{Spectra of wall-pressure fluctuations in boundary layers.  Smooth wall data  at $\Rt=4\,021$  to 11\,064 \citep[grey lines]{fritsch_pressure_2020,fritsch_fluctuating_2022}. Left: log-log form.  Right: pre-multiplied form. }
    \label{fig:loglog}
    \end{figure}
    
\begin{figure}
        \centering
        \includegraphics{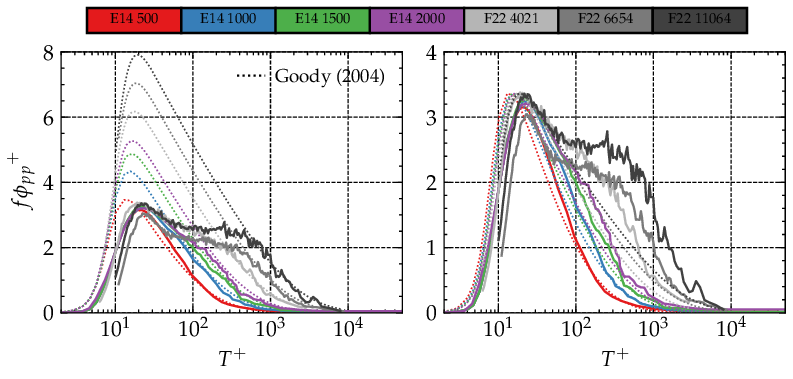}
        \caption{Pre-multiplied spectra of wall-pressure fluctuations in  boundary layers compared with Goody's (2004) model.  Left: model as given.  Right: model normalised so that the peak value is fixed at 3.36.  Highly-resolved LES data from \cite{eitel-amor_simulation_2014} for $\Rt=500$ to 2000,  experimental  data from \cite{fritsch_pressure_2020,fritsch_fluctuating_2022} for $\Rt=4021$  to 11,064.   Model predictions are shown by the dashed lines colour-coded to the data. }
    \label{fig:Goody}
    \end{figure}

    In the Goody model, the $\Rt$-dependence is captured through the timescale ratio between the inner and outer scales, which presents as the growth of the $f^{-1}$ region in $\phi_{pp}$ illustrated in figure~\ref{fig:loglog}. 
    Figure~\ref{fig:Goody} shows the pre-multiplied spectra for the boundary-layer data and the prediction from the Goody model. The model displays a strong Reynolds number dependence that, although very far off at the peak, captures the frequency-dependent growth and decay of the inner- and outer-scales faithfully. {A weakness of the Goody model is that} the matching between the inner- and outer-timescales is done via a \emph{single} modified Lorentzian distribution. A symptom of the fixed shape is that the peak separating the inner- and outer-scale behaviour in the pre-multiplied form has to rise to stay faithful to the gradients of growth and decay of these contributions. The result is a gross mismatch with the data at high Reynolds number (figure \ref{fig:Goody}) although, both the inner and outer growth and decay are captured faithfully.  Even after normalising the {inner} peak to remove Reynolds-number dependence, Goody's model fails to capture the growth in low-frequency (high-$T^+$) energy—the low-frequency portion of the 1-D spectrum.

    The models advanced here build on literature focusing on the increasing energy and length scales of the turbulent structures \citep{smits_high-reynolds_2011} as well as their invariance when scaled with inner and outer variables \citep{wei_properties_2005}. To solve for the pressure, we would need access to the full, 3-D velocity field. Instead, we use measurements of the wall-pressure with the knowledge that {the wall pressure is related to the velocity field through the pressure--Poisson equation, but this connection is non-local and involves multiple source terms; we therefore invoke it only as qualitative motivation.}
    The wall-pressure spectrum $\phi_{pp}$ is constructed in the frequency domain by combining two {spectral components} that broadly represent the contributions to the wall-pressure fluctuations from inner-scaled motions ($g_1$) and outer-scaled motions ($g_2$).

    At sufficiently large $\Rt$, we expect the contributions {represented by $g_1$ to approach an asymptotically inner-scaled form. This expectation is consistent with the observed tendency toward universality at the small-scale end of wall-pressure spectra and with data-driven decompositions of near-wall velocity spectra indicating that, once sufficient scale separation is achieved, the inner contribution collapses in inner scaling (e.g. \citet{baars2020data}). These arguments are heuristic: the mapping from velocity statistics to wall-pressure fluctuations is indirect and non-local, so we treat the saturation of $g_1$ as an empirically guided modelling constraint}. Experimental uncertainty and competing influences from several well-established processes make it difficult to specify the precise $\delta^+$ at which this invariance is achieved. {Nevertheless, established scaling arguments for the velocity field and its decomposition into inner, intermediate and outer contributions provide a useful qualitative guide} \citep{smits_high-reynolds_2011}. Consistent with this, data-driven decompositions show that once a minimal scale separation is present, the near-wall contribution associated with $g_1$ collapses in inner scaling and becomes effectively $\Rt$-invariant. {The second component $g_2$ is expressed in outer time $T^o$ (outer-scaled in this kinematic sense), with its Reynolds-number dependence representing the growing contribution and bandwidth of intermediate/large scales when projected onto $\phi_{pp}(f)$.}

    These contributions are modelled as {spectral} distributions in ${f\phi_{pp}}^+$ that overlap in the $T$ domain. The energy is taken to be a linear summation over the two {spectral components}. That is,
    \begin{equation}
        {f\phi_{pp}}^+=g_1(T^+; \Rt)+g_2(T^o; \Rt).
    \label{eq:combined_model}
    \end{equation}

    Since ${f\phi_{pp}}^+=g_1(T^+;\Rt)+g_2(T^o;\Rt)$ is evaluated at a fixed physical frequency $f=1/T$, the inner- and outer-normalised periods are linked by $T^o=T^+\,(U_e^+/\delta^+)$ with $U_e^+ \equiv U_e/u_\tau$ (here $U_e$ denotes the appropriate outer velocity scale). Hence, both {components} are summed at the same $f$ without invoking a convection velocity; $U_c$ enters only when mapping to $k_x=2\pi f/U_c$. {In the present comparisons, $U_e^+$ is taken directly from the underlying datasets and therefore no additional empirical model is required to relate $T^+$ and $T^o$. For extrapolation to higher $\delta^+$, standard friction-law/log-law correlations may be used to estimate the slow variation of $U_e^+$ with Reynolds number.} At low-$\Rt$, the inner and outer-scales overlap significantly, confusing the distinction between the two {components}. As $\Rt$ increases, the two {components} separate in frequency space, with $g_1$ dominating at high frequencies (low $T^+$) and $g_2$ dominating at low frequencies (high $T^o$).  The models proposed below aim to capture this transition from inner- to outer-scaling as $\Rt$ increases.

    An immediate consequence of \eqref{eq:combined_model} is that we expect to see the appearance of an overlap region at a sufficiently high Reynolds number where $f\phi_{pp}$ is neither a function of $T^+$ nor $T^o$, that is, where $g_1+g_2=constant$,  so that there is an $f^{-1}$ region in $\phi_{pp}$ and a plateau region in $f\phi_{pp}$ (figure~\ref{fig:loglog})\footnote{It is this plateau region that the Goody model fails to capture.}. In wavenumber space, this corresponds to a $k^{-1}$ region, where $k=2\pi f/U_c$ is the streamwise wavenumber and $U_c$ is the convection velocity in this wavenumber range.  This result is in accordance with numerous previous studies (see, for example, \cite{klewicki_statistical_2008}), and we see this overlap region develop with increasing Reynolds number in figure~\ref{fig:loglog}, in both the log-log and pre-multiplied representations. 
    Klewicki {\it et al.\/} also cited \cite{panton_wall_1974} in observing that if $U_c$ is {not} constant, the slope of the $k^{-1}$ region in the wavenumber spectrum is preserved as an $f^{-1}$ region in the corresponding frequency spectrum.

    We offer two versions of this general model.  The first version (model A) {acts as a low-parameter estimation} representing the inner- and outer-components by two log-normal distributions in the pre-multiplied spectrum. Model A follows the approach taken by \citet{gustenyov_model_2025} in representing the spectrum of the streamwise Reynolds stress, providing a family of models for the velocity and pressure spectra. In the second version, we aim to incorporate the known behaviour of the pressure spectra using a modified Lorentzian spectral shape, similar to the approach taken by \citet{goody_empirical_2004}, but with the important separation of the contribution from inner-and outer-{scaled spectral components}. This approach allows the model to incorporate known asymptotic limits on the spectrum, which may therefore allow a more confident extrapolation to very high Reynolds numbers, such as those encountered in realistic engineering examples.  The behaviour of both models is guided by the theoretical understanding of the wall-pressure spectrum and by empirical observations of its scaling with Reynolds number, with the aim of providing a continuous model that captures the transition from inner to outer scaling as $\Rt$ increases.

\section{Model definitions}
\subsection{Model A--Log-Normal}\label{sec:Model A}
    For $g_1$ and $g_2$ in Model A, we will assume that their contributions to the pre-multiplied energy distribution can be modelled using log-normal distributions in $T$.  That is, we propose
    {
    \begin{eqnarray}
        g_{1} & = & A_{1} r_v \exp \left[ -  \left( \frac{ \log{T^+} - \log{\overline{T}^+}}{\log{\sigma_1}} \right)^2 \right]
        \label{g1_model} \\
        g_{2} & = & A_{2} r_v \exp \left[ - \left( \frac{ \log{T^o} - \log{\overline{T}^o}}{\log{\sigma_2}}  \right)^2 \right].
        \label{g2_model}
    \end{eqnarray}
    }
    The energy content  is thus distributed around the (non-dimensional) periods for the inner and outer contributions to the spectrum ($\overline{T}^+, \overline{T}^o$), with the frequency range of the distributions described by ($\sigma_1, \sigma_2$).  $\overline{T}^+$ represents the temporal centre of the energetic contributions from the inner {component}, while $\overline{T}^o$ represents the centre of the contributions from the outer {component}. $\sigma_1, \sigma_2$ represent the width of the distributions, controlling their frequency range.
    A viscous damping term $r_v$ is active for $T^+ \lesssim 15$, defined by the smooth step function
    \begin{equation}
        r_v = \frac{\exp(r_1T^+)}{\exp(r_1\,r_2)+\exp(r_1T^+)};
    \end{equation}
    $r_v$ captures the development of the inner {component}' contribution towards its saturation at high $\delta^+$; this development will vary between boundary layers and internal flows. {Importantly, the invariance of $g_1$ at high $\delta^+$--for pipe and channel flow--is enforced by the asymptote of $r_v$ to 1 at sufficient $\delta^+$.}

    The best fit constants for Model A are provided in {table~\ref{tab:A params}.  The most notable differences among the three flow types are in the Reynolds number dependence of $A_1$ (which becomes negligible at high Reynolds number) and the location of the peaks as given by $\overline{T^+}$ and $\overline{T^o}$}.

    \begin{table}
    \centering
    \resizebox{\columnwidth}{!}
    {\begin{tabular}{lcccccccccc} \small
        & & & & & & & & & \\[-5mm]
        & &  \ $A_1$ \ & \ $\sigma_1$ \ &  \ $\overline{T}^+$ \ &  &  \   $A_2$ \  & \  $\sigma_2$ \  & \ $\overline{T}^o$ \  & \ $r_{1}$ \ & \ $r_{2}$ \  \\
        & & & & & & & & & \\[-1.5mm]
        Boundary layer & & $2.20 $  & 3.90  & 20 & & $1.40 (\log{\Rt} -2.2)$   & 8.18 & 0.82  & 0.50 & 7 \\
        Pipe & &   $2.90 (1-1000/\Rt) $  & 4.30  & 20 & & $0.91 (\log{\Rt} -2.2)$   & 10.0 & 0.18  &  0.50 & 7 \\
        Channel &  &  $2.10 (1-100/\Rt) $  & 4.40  & 12 & & $0.90 (\log{\Rt} -2.2)$   & 10.0 & 0.60  & 0.50 & 3 \\[1mm]
        & & & & & & & & &  \\[-3mm]
    \end{tabular}}
    \caption{Model A best-fit constants.}
    \label{tab:A params}
    \end{table}

\subsection{Model B--Modified Lorentzian}
    In Model B, our aim is to develop the model to capture known behaviour of the wall-pressure spectrum using a modified Lorentzian spectral shape, similar to the approach of \citet{goody_empirical_2004}. We focus on the pipe-flow data from \citet{dacome_scaling_2025} and use scaling arguments introduced above to extend the model to the boundary-layer data from \citet{fritsch_pressure_2020,fritsch_fluctuating_2022}. A key difference from \eqref{eq:combined_model} is that $g_1$ is no longer a function of $\Rt$ as the data in \citet{dacome_scaling_2025} has an inner component that is fully developed. We propose that the general form of the pre-multiplied wall-pressure spectrum should be given by
    \begin{equation}
        g_i
        = A\,2^{r}
        \Bigl(\frac{T_b}{T}\Bigr)^{p_{\mathrm{low}}}
        \Bigl[1 + \Bigl(\frac{T_b}{T}\Bigr)^{q}\Bigr]^{-r}\,,
    \end{equation}
    which for $T_b/T \ll 1$ reduces to
    \begin{equation}
        g_i \sim A\,2^{r}
        \Bigl(\frac{T_b}{T}\Bigr)^{p_{\mathrm{low}}}\, \Longrightarrow
        \phi_{pp}^+ \propto f^{p_{\mathrm{low}}-1}\,,\,\frac{\mathrm{d}\ln\phi_{pp}^+}{\mathrm{d}\ln f}
        \sim p_{\mathrm{low}} - 1
    \label{eq:general low}
    \end{equation}
    {and $T_b$ is analogous to $\overline{T}^+$ and $\overline{T}^o$ defined in \S~\ref{sec:Model A}. Note that exponents here refer to the premultiplied spectrum $f\phi_{pp}^+$; consequently, the power-law exponent of $\phi_{pp}^+$ is lower by one.}
    Similarly, for $T_b/T \gg 1$,
    \begin{equation}
        g_i \sim A\,2^{r}\,\Bigl(\frac{T_b}{T}\Bigr)^{p_{\mathrm{low}}}\,
            \Bigl(\frac{T_b}{T}\Bigr)^{- q\,r}\,\Longrightarrow
        \phi_{pp}^+ \propto f^{p_{\mathrm{low}} - q\,r - 1}\,, \,\frac{\mathrm{d}\ln\phi_{pp}^+}{\mathrm{d}\ln f}
        \sim p_{\mathrm{low}} - q\,r - 1 \,.
    \label{eq:general high}
    \end{equation}
    To characterise the sharpness of the transition at $T = T_b$, define $\epsilon = T_b/T$ and $f(\epsilon) = \bigl[1 + \epsilon^{q}\bigr]^{-r}$ so
    \begin{equation}
        \frac{\mathrm{d}\ln f}{\mathrm{d}\ln\epsilon}
        = -\,r\,\frac{q\,\epsilon^{q}}{1 + \epsilon^{q}}\,, \quad \Delta(\log\epsilon) \approx \frac{2}{q}\,.
    \label{eq:sharpness_slope}
    \end{equation}
    Thus $q$ directly controls the transition sharpness; larger $q$ yields a narrower region between the low- and high-frequency asymptotes.

    Finally, we set
    \begin{equation}
        r = \frac{p_{\mathrm{low}} - p_{\mathrm{high}}}{q}\,,
    \end{equation}
    so that as $T_b/T\to\infty$, the high-frequency exponent becomes $p_{\mathrm{high}} = p_{\mathrm{low}} - q\,r$.

\subsubsection{Inner-scale component}
    The explicit form of the {inner function} is given with
    \begin{equation}
        g_1
        = A^{\text{in}}\,2^{r^{\text{in}}}
        \Bigl(\frac{T_{b^{\text{in}}}^+}{T^+}\Bigr)^{p_{\mathrm{low}}^{\text{in}}}
        \Bigl[1 + \Bigl(\frac{T_{b^{\text{in}}}^+}{T^+}\Bigr)^{q^{\text{in}}}\Bigr]^{-r^{\text{in}}}\,,
    \end{equation}
    where the parameters are defined as before and $(\cdot)^{\text{in}}$ indicates the parameter associated with the inner {peak}. The break period $T_{b^{\text{in}}}^+$ is defined in inner units, and the amplitude $A^{\text{in}}$ is a constant that sets the magnitude of the inner {peak}'s plateau.
    
    Following Townsend's attached-eddy model, which predicts $\phi_{pp}(f) \sim f^0$ as $f \to 0$ for smooth-wall turbulent flows \citep{townsend_structure_1976}, we select $$p^{\text{in}}_{\mathrm{low}} = 1\,.$$

    For the high-frequency decay, we are guided by the classical rapid-decay theories: \citet{kraichnan_pressure_1956} suggested a steep spectrum $\phi_{pp} \sim f^{-5}$ at very high frequencies. In our formulation, a high-frequency decay of $\Phi_{pp,\mathrm{in}}^+ \sim (f^+)^{-5}$ corresponds to
    $$p^{\text{in}}_{\mathrm{high}} = -6\,.$$

    Prior studies noted that in an intermediate range around the {inner} peak, the wall-pressure spectrum often follows $\phi_{pp} \sim f^{-1}$ \citep{bradshaw_inactive_1967,panton_wall_1974,blake_mechanics_1986}. To incorporate this overlap scaling, we adjust $r^{\text{in}}$ such that the slope at $f \approx f_{b^{\text{in}}}^+$ is $-1$. For a symmetric Lorentzian ($q^{\text{in}} = 2$), this condition is approximately met by $r^{\text{in}} \approx 2$. We therefore take $r^{\text{in}} = 2$ as a convenient choice that yields an overlap slope of order $-1$ (and a slightly steeper ultimate decay, closer to $f^{-6}$, at the highest frequencies). It should be noted that there is some debate in the literature regarding the exact value of the transition slope, \citet{goody_empirical_2004,klewicki_statistical_2008} suggesting values closer to $0.8$.

    The break period $T_{b^{\text{in}}}^+$ is set based on the frequency at which the near-wall (inner) spectral contribution begins to roll off. Using experimental data for smooth-wall turbulence, we choose $f_{b^{\text{in}}}^+ = 0.1$ following the observations of \citet{morrison_interaction_2007}, who identified a spectral inflection (associated with the buffer-layer peak) around that value in inner units\footnote{This is an observation for a boundary-layer, but through the two different modelling approaches we find the break period remains consistent across pipes and boundary-layers}.

    Finally, the amplitude $A^{\text{in}}$ is tuned by matching the variance of the inner-model to the variance of channel data at $\Rt \approx 1\,000 $ (figure~\ref{variance}). The rationale behind this is that the {inner} peak is almost fully developed at this $\Rt$ and there is only a small influence from the outer-scale energy. This yields a value of $A^{\text{in}} \approx 1.6$. A summary of the inner parameters is given in Table~\ref{tab:B params}.

    \begin{table}
        \centering
        \begin{tabular}{l|cccccccc}
            \ Inner {function} \ & \ $A^{\text{in}}$ \ & \ $T_{b^{\text{in}}}^+$ \ & \ $p^{\text{in}}_{\mathrm{low}}$ \ & \ $p^{\text{in}}_{\mathrm{high}}$ \ & \ $q^{\text{in}}$  \\
             &  &  &  &  & \\[-2mm]
            & 1.6 & 10 & 1 & -6 & 3.5  \\
            \hline
            \ Outer {function} \ & \ $T_b^{\text{out}}$ \ & \ $p^{\text{out}}_{\mathrm{low}}$ \ & \ $a_A$ & $b_A$ \ & \ $a_p$ \ & \ $b_p$ \ & \ $a_q$ \ & $b_q$ \ \\
            &  &  &  &  &  &  &  \\[-2mm]            
            & 1.5 & 3 & 2.39 & 3.55 & -0.22 & 3.58 & -1.09 & 3.78 \\
        \end{tabular}
        \caption{Model B constants.}
        \label{tab:B params}
    \end{table}

\subsubsection{Outer-scale component}

    We now formulate the outer-scale contribution in an analogous manner. Explicitly
    \begin{equation}
        g_2
        = A^{\text{out}}\,2^{r^{\text{out}}}
        \Bigl(\frac{T_b^o}{T^o}\Bigr)^{p^\text{out}_{\mathrm{low}}}
        \Bigl[1 + \Bigl(\frac{T_b^o}{T^o}\Bigr)^{q^\text{out}}\Bigr]^{-r^\text{out}}\,,
    \label{eq:outer_form}
    \end{equation}
    where all parameters ($A^{\text{out}}$, $p^{\text{out}}_{\mathrm{low}}$, $T_b^{\text{out}}$, $q^{\text{out}}$, $r^{\text{out}}$) pertain to the outer component.

    We set $p^{\text{out}}_{\mathrm{low}} = 3$ consistent with the notion of the outer pressure field being generated by a relatively smooth (slowly evolving) process \citep{cramer_stationary_2013}. It implies that at the lowest frequencies the outer pressure fluctuations are significantly attenuated (a $\sim f^2$ spectral rise from the origin, as opposed to a flat spectrum).

    We associate the outer break period, $T_b^{\text{out}}$, with the characteristic turnover frequency of the largest attached eddies in the flow. This is related to the convective timescale of outer structures, on the order of $\delta/U_\delta$. We choose $T_b^{\text{out}}$ such that
    \begin{equation}
        T_b^{\text{out}} = \frac32
    \end{equation}
    in outer units, meaning that $T_b^{\text{out}} = 3/2$ of a cycle per outer flow time, similar to the arguments presented by \citet{jacobi_interactions_2021}. This choice is guided by prior observations of the convection speed of energetic outer-scale motions \citep{morrison_interaction_2007,mckeon_critical-layer_2010,jacobi_interactions_2021}, which indicate that the spectral peak associated with large-scale structures occurs at a fraction of the free-stream velocity (for boundary layers) or centreline velocity (for pipes).

    A summary of the outer model parameters is given in Table~\ref{tab:B params}. The values of $A^{\text{out}}$, $p^{\text{out}}_{\mathrm{high}}$, and $q^{\text{out}}$ are determined from the training procedure described in the next section.

    \subsubsection{Fitted parameters for the outer component}

        The outer parameters $A^{\text{out}}$, $p^{\text{out}}_{\mathrm{high}}$, and $q^{\text{out}}$ {are left free to empirically fit to the available data and} aim to capture the $\Rt$ dependent behaviour of the pressure spectra. At low $\Rt$, outer structures are weak relative to the {inner-scaled peak} implying a low amplitude, steep decay, and rapid roll-off. At high $\Rt$, outer structures become stronger and populate a broader frequency range, meaning the amplitude is larger, decays more slowly, and the roll-off is less steep. To incorporate this $\Rt$ dependence, we allow $A^{\text{out}}$, $p^{\text{out}}_{\mathrm{high}}$, and $q^{\text{out}}$ to vary with $\Rt$. Specifically, we choose a logistic (sigmoidal) form for these dependencies, which ensures smooth transition between asymptotic values at low and high $\Rt$:
        \begin{subequations}
            \begin{align}
                A^{\text{out}}(\Rt) &= \frac{1}{\,1 + \exp\bigl[a_A\!(b_A - \log\Rt)\bigr]}\,, \\
                p^{\text{out}}_{\mathrm{high}}(\Rt) &= -2 +\frac{1.5}{\,1 + \exp\bigl[a_p\!(b_p - \log\Rt)\bigr]}\,, \\
                q^{\text{out}}(\Rt) &= 0.2+\frac{0.6}{\,1 + \exp\bigl[a_q\!(b_q - \log\Rt)\bigr]}\,,
            \end{align}
        \label{eq:outer sigmoids}
        \end{subequations}
        where $a_A, b_A, a_p, b_p, a_q, b_q$ are constants determined from data fits. The asymptotes are chosen to embed the observation that
        \begin{equation}
            \int^\infty_0 g_2 \, d\!\log\!f \mapsto \approx 0 \quad \text{as} \quad \Rt \to 1\,000
        \end{equation}
        \\
        and to ensure stability as $\Rt\to\infty$.

    \subsubsection{{Fitting} procedure and results}

        The model is {optimised} on the wall-pressure spectra from \citet{dacome_scaling_2025} at $\Rt \in [4\,794,47\,015]$. {We note that the highest-$\delta^+$ pipe spectra require facility-specific corrections (e.g. Helmholtz resonance and background-noise rejection), and such corrections are never perfect for a 1-D frequency spectrum (Appendix \ref{sec:uncertainties}). The imposed decay rates in this B form reduce concerns about overfitting to the extreme low- and high-frequency ends of the calibration spectra. However, the fitted  parameters may carry some systematic uncertainty. This should be kept in mind when comparing the calibrated models to other/future measurements}.
        The fitting procedure involves minimising the loss function, which is defined as the sum of the squared differences between the modelled and measured spectra, as well as a weighted difference between the modelled and theoretical variance proposed by \citet{lee_direct_2015}
        \begin{equation}
            \langle p^{2}_w\rangle^+=2.24 \ln{\Rt}-9.18 \,.
        \label{eq:var_fit}
        \end{equation}
        Mathematically, the loss function is defined as
        \begin{equation}
            \mathcal{L} = \sum_{\omega}\bigl[f\phi_{pp}^{\mathrm{model}}-f\phi_{pp}^{\mathrm{data}}\bigr]^2
      + \bigl[\langle p_w^{2}\rangle^+_{\mathrm{model}}
             -\langle p_w^{2}\rangle^+_{\mathrm{LM15}}\bigr]^{0.02} \, .
        \label{eq:variance tuned loss function}
        \end{equation}
        The parameters are optimised using a Nelder-Mead minimisation. The optimised parameter values are reported in table~\ref{tab:B params}. {The variance term in eq.~\eqref{eq:variance tuned loss function} encourages consistency with the established channel-flow trend of eq.~\eqref{eq:var_fit}--although the contribution is limited by the small power coefficient--and therefore the variance scaling aims to be \emph{captured} rather than independently predicted by the model.}

\section{Results}
    \subsection{Model A--Log-Normal}
\begin{figure}
            \centering
            \includegraphics{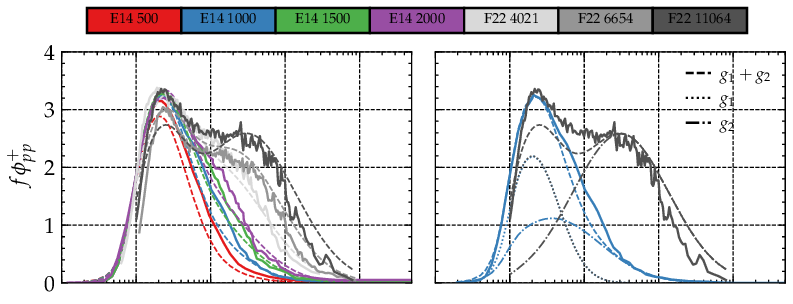}
            \includegraphics{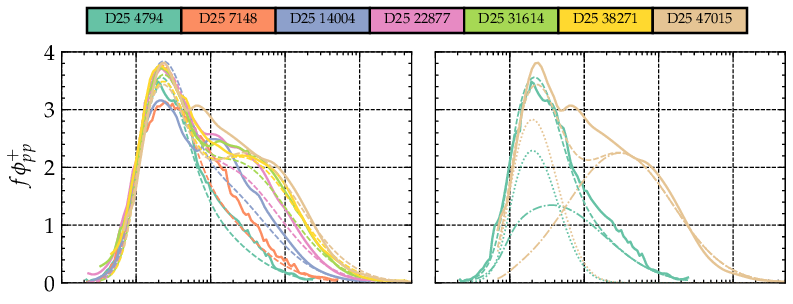}
            \includegraphics{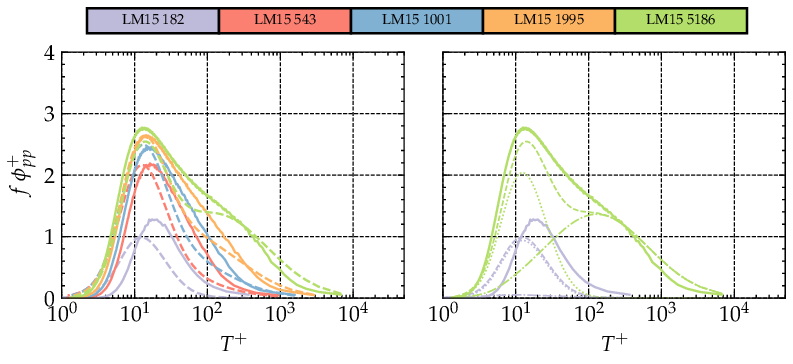}
        \caption{Comparison of Model A with data shown in figure~\ref{fig:data}. Model constants listed in table~\ref{tab:A params}.   Top row: boundary layers.  Left: all $\Rt$.  Right: $\Rt= 1\,000$, 11\,064, showing $g_1$ and $g_2$ ($g_1$ is identical for these $\Rt$). Middle row: pipes.  Left: all $\Rt$.  Right: {$\Rt= 4\,794$, 47\,015}, showing  $g_1$ and $g_2$.  Bottom row: channels.  Left: all $\Rt$.  Right: $\Rt= 550$, 5\,200 showing  $g_1$ and $g_2$.  }
        \label{model}
        \end{figure}

        Model A comparisons with the data for boundary layers, pipes, and channels are shown in figure~\ref{model} (left column). Best fits to the data were obtained using the constants listed in table~\ref{tab:A params}.  For all three flows, over the entire Reynolds number ranges covered by the data, the model gives excellent agreement with the data.  In the right column, two cases have been picked out for each flow type, separated by about a factor of 10 in Reynolds number.  These examples illustrate how well the model reproduces the spectrum at all Reynolds numbers explored here.  In addition, we see how $g_1$ and $g_2$ contribute to the total energy content, how they display significant overlap in $T^+$ over the full Reynold number range, and how the {inner peak} evolves with Reynolds number. 

        The amplitudes $A_1$ for pipes and channels are Reynolds number dependent, but only at the lower Reynolds numbers.  The amplitudes $A_2$ for all three flow types depend on Reynolds number in an identical manner, with a fixed  offset of 2.2, corresponding to a Reynolds number of  180.  The values of $\overline{T}^+$ and $\overline{T}^o$ are more or less as expected from our earlier discussion, with the exception of  $\overline{T}^o$ for pipes, which is considerably smaller than the values for boundary layers and channels.  This suggests a comparatively slower growth of the low-frequency contribution in pipes—a point not emphasised in prior work.

    \subsection{Model B--Modified Lorentzian}

        \begin{figure}
            \centering
            \includegraphics{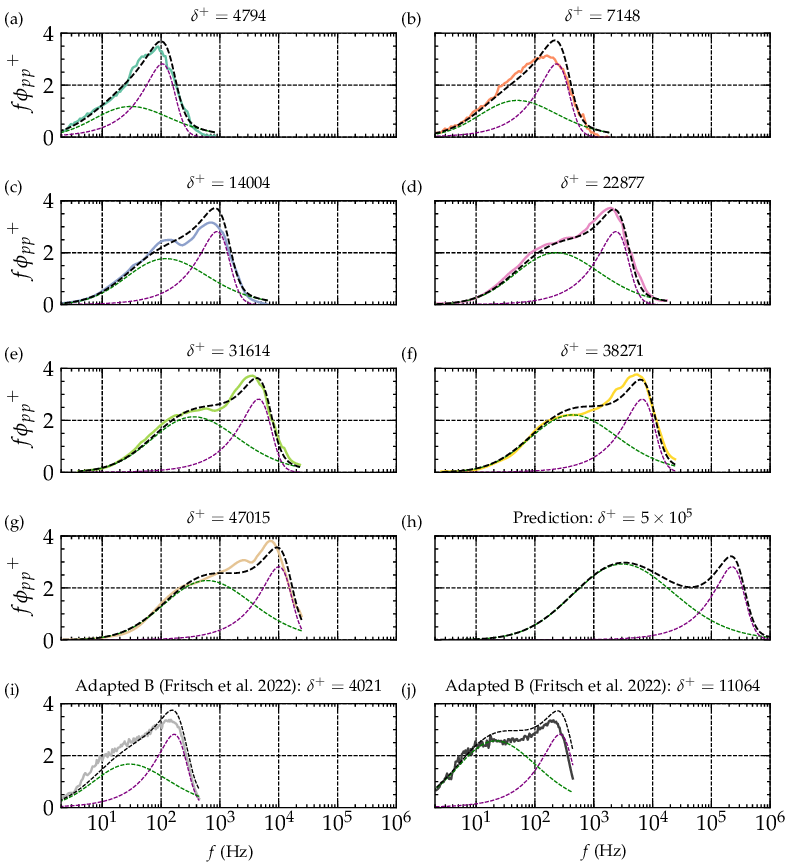}
            \caption{The modelled vs measured wall-pressure spectra at the range of Reynolds numbers measured in \citet{dacome_scaling_2025} (a-g). {The solid lines are the measured spectra, the dashed lines are the modelled spectra. The purple dashed line is $g_1$ for model B, and the green dashed line is $g_2$.} In panel (h), {illustrative extrapolation} of the spectrum at $\Rt=5\times 10^5$. Panels (i,j) correspond to the data and the adapted model B predictions at the labelled $\Rt$ values.}
        \label{fig:expanded view}
        \end{figure}

        With similar success, Model B matches well with the inner and outer components summed to reconstruct the original wall-pressure spectrum given by \eqref{eq:combined_model}. The expanded view in figure \ref{fig:expanded view} shows the modelled vs measured wall-pressure spectra at the range of Reynolds numbers measured in \citet{dacome_scaling_2025}. By design, the inner-scaled contribution remains invariant with $\Rt$, and the outer-scaled contribution varies with $\Rt$.  In figure \ref{fig:expanded view}h, {we show an illustrative extrapolation of Model B to $\delta^+ = 5\times 10^5$ to demonstrate the internal consistency of the continuous asymptotic formulation. This extrapolation does not constitute independent validation and should be interpreted cautiously given the finite calibration range and the experimental uncertainties at the spectral extremes (Appendix \ref{sec:uncertainties})}. The inner-peak in the predicted spectrum has moved outside the range of human hearing here and the outer-peak centres around 2 000 Hz and contains the majority of the energy.

        Model B was primarily developed for turbulent pipe flow, but it can be extended to boundary-layer flows by adjusting some chosen parameters. Namely, we change the outer break period to $T_b^{\text{out}} = 3.45$, consistent with the longer outer-scaled structures observed by \citet{lee_comparison_2013} in boundary-layer flows. The amplitude $A^{\text{out}}$ is also increased by a factor of $1.56$ to account for the different scaling of the wall-pressure spectrum in boundary layers. The inner component remains unchanged, as the near-wall pressure fluctuations are expected to be similar in both pipe and boundary-layer flows.  The resulting model for boundary-layer flows is tested against \citet{fritsch_fluctuating_2022}.  Although this is not a true prediction, as the coefficients are tuned using observed trends, good agreement is obtained for both the low-$\Rt=4\,021$ and high-$\Rt=11\,064$ cases.
    
    \subsection{Variance}\label{sec:A var}

    Turbulent wall-pressure fluctuations are known to intensify with increasing Reynolds number. Both experimental and numerical studies have observed that the wall-pressure variance, $\langle p^{\prime 2}_w \rangle^+ = \langle p^{\prime 2}_w \rangle /\tau_w^2$,  grows approximately logarithmically with the friction Reynolds number, $\Rt$ \citep{farabee_spectral_1991,panton_correlation_2017}. This behaviour is consistent with Townsend's (1951) attached-eddy hypothesis, \nocite{townsend_structure_1951} which postulates that as $\Rt$ increases a broader range of self-similar eddies contributes to the pressure field, producing a $k_x^{-1}$ spectral region whose integration leads to the scaling $\langle p'^2_w \rangle^+ \propto \ln(\Rt)$. 
    For instance, the boundary-layer experiments of \citet{farabee_spectral_1991}, in the range $\Rt \approx 10^3$-$2\times10^3$, clearly demonstrated a rise in $\langle p^{\prime 2}_w\rangle^+$ with Reynolds number, which they attributed to an expanding $f^{-1}$ range in the pressure spectrum.  Despite the evident growth of the {low-frequency} range with Reynolds number, and the consequent departure from $f^{-1}$ scaling, the logarithmic dependence appears to be quite robust, even at very high Reynolds numbers \citep{klewicki_statistical_2008,panton_correlation_2017}.

\begin{figure}
        \centering
        \includegraphics{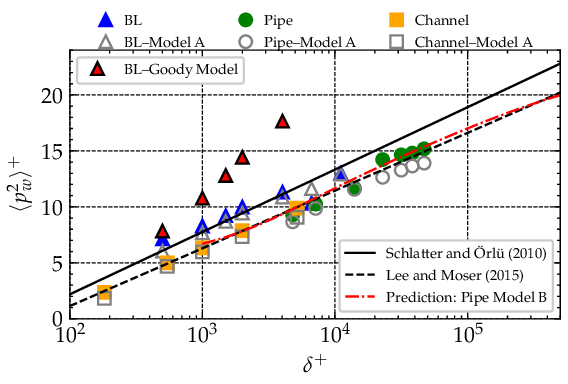}
        \caption{Variance of wall-pressure fluctuations.  The solid markers are the data described in figure~\ref{fig:data}, where: \textcolor{blue}{$\blacktriangle$} is the boundary-layer data, \textcolor{green}{$\bullet$} the pipe, and \textcolor{orange}{$\blacksquare$} the channel.  The Goody-model variance is \protect\goodymarker.  Model A is shown with grey open symbols matching the data.  Model B over $\log \Rt\!\in[3,5]$ is denoted by $\textcolor{red}{-\cdot}$.  For comparison, the empirical relations are $-$ for the boundary-layer correlation $\langle p^{\prime 2}_w\rangle^+=2.42\ln\Rt-8.96$ \cite{schlatter_assessment_2010} and $--$ for the channel correlation $\langle p^{\prime 2}_w\rangle^+=2.24\ln\Rt-9.18$ \cite{lee_direct_2015}.}
        \label{variance}
    \end{figure}

    The variance $\langle p'^2_w \rangle^+$ is found by integrating the spectra over all frequencies.  The results for both models and the underlying data are shown in  figure~\ref{variance}.  As expected from the good agreement between the model spectra and the data, the variances calculated for the model and data agree very well.  Furthermore, the boundary layer results for Model A agree well with the correlation proposed for boundary layers by \cite{schlatter_assessment_2010} ($\langle p^{\prime 2}_w\rangle^+=2.42 \ln{\Rt}-8.96$), and the pipe and channel results agree well with the correlation proposed for channels by \cite{lee_direct_2015} \eqref{eq:var_fit}, with the continuous form of Model B also aligning with \eqref{eq:var_fit} {by design (cf.~\eqref{eq:variance tuned loss function}) and extending in its continuous predictive form to $\Rt=5\times 10^5$.}  {The similarity across the channel and pipe variance matches well with the findings of \citet{yu_reynolds_2022,wei_scaling_2025}}. The Goody model shows a major disagreement with these other trends, significantly overpredicting the variance.

    We can also relate the wall pressure variance to the wall shear stress variance by {combining} the correlation obtained by \cite{samie_fully_2018} for $\langle u^{\prime 2}_p\rangle^+$, the maximum value of the inner peak in the streamwise Reynolds stress, 
    \begin{equation}
        \langle u^{\prime 2}_p\rangle^+ = \frac{\langle u^{\prime 2}_p\rangle}{u_\tau^2} = 3.54 + 0.646 \, \ln{\Rt}.
    \end{equation}
    with, 
    \begin{equation}
        \langle u^{\prime 2}_p\rangle^+ \approx 46 \langle \tau^{\prime 2}_w\rangle^+
    \end{equation}
    \citep{smits_reynolds_2021,chen_reynolds_2021}.  Then, by using the correlation for $\langle u^{\prime 2}_p\rangle^+$ proposed by \cite{schlatter_assessment_2010},
    \begin{equation}
        \frac{\langle p^{\prime 2}_w\rangle^+}{\langle u^{\prime 2}_p\rangle^+} \approx \frac{\langle p^{\prime 2}_w\rangle^+}{46 \langle \tau^{\prime 2}_w\rangle^+}  = \frac{2.42 \, \ln{\Rt}-8.96}{3.54 + 0.646 \, \ln{\Rt}},
    \end{equation}

    we obtain a {tentative} connection between the variances in wall pressure and wall shear stress--in addition to connecting both with the magnitude of the inner peak in $\overline{u^2}$. This suggests a previously unreported correlation between wall-pressure and wall-shear stress variances based on reported correlations using the same variables. {This construction is heuristic and should not be interpreted as a derivation from the governing equations or as an established causal connection; its validity and universality (including the assumed linkage between the inner peak and wall-shear behaviour) remain an active topic of discussion in the literature (e.g. \citet{marusic_scaling_2017}). Direct, simultaneous measurements of wall-pressure and wall-shear-stress fluctuations at high Reynolds number would be required to assess whether such a relationship holds quantitatively.}

\section{Discussion \& Consequences}

    We have shown that it is possible to model the energy content of the wall pressure signal using two functions: an inner-scaled function $g_1$ and an outer-scaled function $g_2$.  
    Both models proposed here reproduce the pre-multiplied spectra and the variances for boundary layers and pipes, with model A extending down to the channel flow data.
    
    \begin{figure}
        \centering
        \includegraphics{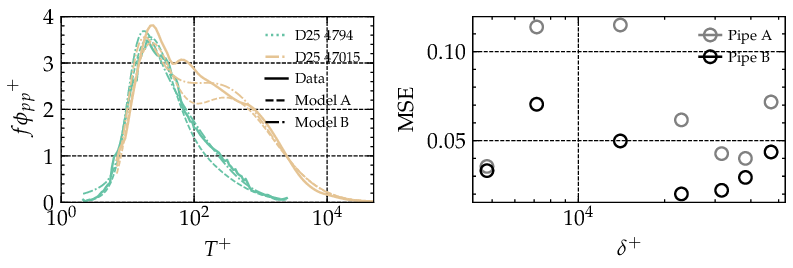}
        \caption{Comparing Model A and Model B with pipe flow data. Left: data from \citet{dacome_scaling_2025} at $\Rt = 4\,794$ to $47\,015$. Right: Mean Squared Error (MSE) for Model A and Model B.}
    \label{fig:pipe_model_comp}
    \end{figure}

    Models A and B are compared in figure~\ref{fig:pipe_model_comp} for pipe flow at two Reynolds numbers, $\Rt = 4\,794$ and $\Rt = 47\,015$.  The left panel shows the spectra along with the modelled forms, and the right panel shows the mean squared error (MSE) between the two models across the frequency range. The MSE is calculated as
    \begin{subequations}
        \begin{equation}
            \mathrm{MSE_A} = \frac{1}{N}\sum_{i=1}^{N}\bigl(\phi_{pp}^{\mathrm{Model A}}(f_i) - f\phi_{pp}(f_i)^+\bigr)^2\,,
        \end{equation}
        \begin{equation}
            \mathrm{MSE_B} = \frac{1}{N}\sum_{i=1}^{N}\bigl(\phi_{pp}^{\mathrm{Model B}}(f_i) - f\phi_{pp}(f_i)^+\bigr)^2\,,
        \end{equation}
    \end{subequations}
    where $N$ is the number of frequency points and $f_i$ are the discrete frequency values. Both models show a good agreement with the data, with model B generally performing better despite the extra asymptotic constraints. {The close agreement between the models is a feature as we believe they serve a complementary purpose: Model A for compact reproduction with minimal inputs; Model B when one wishes to enforce asymptotic exponents and smooth $\delta^+$-dependence (eqs.~\eqref{eq:outer sigmoids}a-c)}.
    
    For both models, the function $g_1$ models the {inner-scaled peak}, and it has a characteristic time constant $\overline{T}^+ = 12$-20 (see table~\ref{tab:A params} and \ref{tab:B params}).  The convection velocity for this {{inner-scaled peak}} is often taken to be $U_c^+ =10$-12 \citep{ghaemi_piv-based_2012, dacome_scaling_2025}, which corresponds approximately to the mean velocity at the location of the inner peak in the streamwise Reynolds stress $\overline{u_p^2}^+$ (located at $y^+ \approx 15$). As mentioned (cf. \S1), the convection velocity is wavenumber dependent, our use of constant values helps to give insight without being prescriptive. This time constant $\overline{T}^+$  is an order of magnitude smaller than the average period of the near-wall bursting events, which is about 100 \citep{metzger_scaling_2010}. The corresponding wavelength $\lambda_{x1} = U_c\overline{T_1}$, so that $\lambda_{x1}^+ = 10\overline{T}^+ = O(100)$, and it matches the characteristic spacing between the near-wall streaks \citep{smith_characteristics_1983}.
    {The lower $T^+$ of the inner-peak observed in channel flow compared with pipe and boundary-layer flows likely arises from the length-scale dependence of the convection velocity. The channel DNS data of \citet{lee_direct_2015} are converted to frequency space using a fixed $U_c^+=10$; however, \citet{del_alamo_estimation_2009,damani_measurement_2025} show that high-$k_x$, short-$T$ motions, and subsequent pressure fluctuations, convect more slowly, which would artificially shift the inner peak to lower $T^+$. For reference, the peak in the time-resolved spectra of \citet{anantharamu_analysis_2020} occurs at $T^+\approx 18$, consistent with the pipe and boundary-layer data. }

    As a complement to $g_1$, the function $g_2$ models the {outer-scaled peak}, with a characteristic time constant $\overline{T}^o = 0.2$-0.8 (see table~\ref{tab:A params}).  The convection velocity connected with this {{outer-scaled peak}} is $U_c=0.7U_e$ \citep{damani_characteristics_2024}\footnote{Many other values have been proposed, ranging from $0.6U_e$ \citep{chase_turbulent_1982} to $0.819U_e$ \citep{hu_aeroacoustics_2002}.}, which corresponds to the mean velocity at a location near the outer part of the logarithmic region.  The matching wavelength $\lambda_{x2}^o = U_c/U_e\overline{T^o}$, so that the streamwise wavelength $\lambda_{x2}^o = 0.7 \overline{T}^o \approx 0.14$-0.56, and the matching wavenumber is $k_x^o \approx 11$-45.  The evidence, therefore, suggests that the {outer-scaled peak} is associated with motions typical of the inertial-layer; these are considerably smaller than the size of the LSM and VLSM. It would be of interest to relate $g_2$ more explicitly to the attached-eddy picture, but this would require a more detailed study to disentangle the wall-pressure imprint of different eddy populations in the projected spectrum.

    In this paper, we have focused on canonical flow cases,  demonstrating that our simple  models can give an accurate representation of the wall-pressure spectrum over a wide range of Reynolds numbers. However, the models are not expected to be valid for flows with significant pressure gradients, compressibility effects, or roughness, as these conditions can significantly alter the scaling of the wall-pressure spectrum. Nevertheless, following on from the approach taken by \citet{gustenyov_model_2025}, we believe that our simple modelling approach can be extended to incorporate these variations in flow physics by adjusting the components accordingly. {For example, we might model roughness by changing $g_1$ and pressure gradient by altering $g_2$}.  In this way, we hope to learn more about the physical underpinning the influence of our two basic {spectral components} on the wall-pressure spectrum.

    The model faithfully reproduces the temporal wall-pressure spectra and--as \citet{Damani_Butt_Totten_Devenport_Lowe_2025} demonstrate--the spatial relationship is required to complete the picture. Ongoing work looks at understanding the physical mechanisms responsible for the logarithmic growth of $\phi_{pp}$ as part of the Shear stress and Propagating Pressure measurements in High Reynolds number Experiments (SAPPHiRE) campaign.

\FloatBarrier

\backsection[Acknowledgements]{We would like to thank the authors of the original datasets for making their data openly available, and for their efforts in collecting high-quality data at high Reynolds numbers. We also thank Dr. Thomas Jaroslawski, Prof. Ivan Marusic, Dr. Vijaya Gudla for their helpful comments in the development of this work.}

\backsection[Funding]{The partial support of DARPA under award \# HR0011-24-9-0465 is gratefully acknowledged.}

\backsection[Declaration of interests]{The authors report no conflict of interest.}

\backsection[Data availability statement]{The data that support the findings of this study are openly available in repositories associated with \citet{fritsch_pressure_2020,fritsch_fluctuating_2022} and \citet{lee_direct_2015}. The pipe flow data were reconstructed from \citep{dacome_scaling_2025}. Upon request, we are happy to share the model code used to generate the results in this paper.}

\backsection[Author ORCIDs]{J.M.O. Massey, https://orcid.org/0000-0002-2893-955X; A. J. Smits, https://orcid.org/0000-0002-3883-8648; B.J. McKeon, https://orcid.org/0000-0003-4220-1583.}

\appendix

\section{Observations on the data}\label{sec:uncertainties}

\begin{figure}
        \centering
        \includegraphics[width=0.48\textwidth]{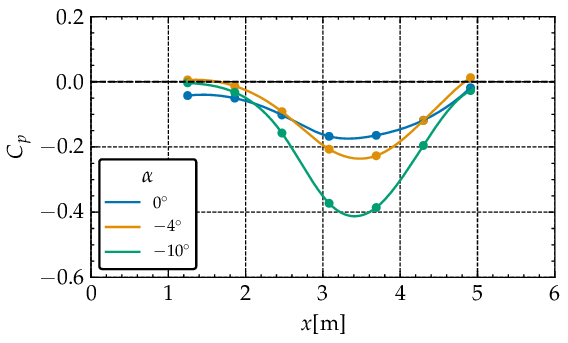} \\
        \hspace{-4mm}
        \includegraphics[width=0.9\textwidth]{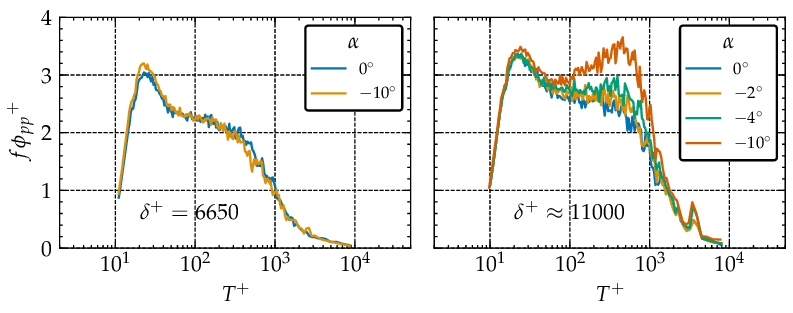}
        \caption{Data from UVA experiments on the tunnel wall for different airfoil angles of attack ($\alpha$) at 58 m/s \citep{fritsch_pressure_2020,fritsch_fluctuating_2022}.  (a) Pressure coefficient distributions.  (b)  Pre-multiplied spectra of wall-pressure fluctuations at $x=1.25$m ($\Rt = 6650$).   (c)  Pre-multiplied spectra of wall-pressure fluctuations at $x=4.91$m ($\Rt \approx 11000$).   }
    \label{UVA_pg} 
    \end{figure}

    There are three possible issues surrounding the quality and completeness of the data used in the modelling.  First, the highest Reynolds number boundary layer profile ($\Rt=11\,064$) was obtained downstream of a mild pressure gradient history imposed on the tunnel wall by the presence of an airfoil in the freestream \citep{fritsch_pressure_2020,fritsch_fluctuating_2022}.  In figure~\ref{UVA_pg}a, we show the pressure coefficient distributions measured when the airfoil was placed at angles of attack $\alpha= 0^o$, -4$^o$ and -10$^o$.  Figure~\ref{UVA_pg}b demonstrates that the presence of the airfoil has little effect on the wall pressure spectrum at the most upstream station ($x=1.95$m), even at $\alpha=-10^o$.  Figure~\ref{UVA_pg}c  similarly demonstrates that for $\alpha<4^o$ the presence of the airfoil has little effect on the wall pressure spectrum at the most downstream station ($x=4.91$m) where the $\Rt=11\,064$ profile was measured.   We propose, therefore, that all the experimental data on boundary layers shown in figure~\ref{fig:data} are free of any significant effects of pressure gradient.
    
    Second, the boundary layer LES and channel flow DNS were obtained in a limited domain which may affect the resolution of the largest outer-scale motions.  For the boundary layer data, the computation is for a spatially-evolving flow, and so the principal limitation on resolving the wall pressure spectrum is the maximum averaging time.  Although \cite{eitel-amor_simulation_2014} do not specify the sampling time, the earlier work by \cite{schlatter_assessment_2010} indicate values of 50\,000 viscous time units, or 36 outer time units at $Re_\theta = 4300$ ($\Rt =1370$).  
    From figure~\ref{fig:data}, we see that this appears to be sufficient to resolve the complete spectrum for each Reynolds number.
    For the channel flow data, the domain size could be a limitation, but for this computation it was $8\pi \delta$, corresponding to $T^+=660$ at $\Rt=550$ and $T^+=4900$ at $\Rt=5200$.  From figure~\ref{fig:data}, we see that this appears to be sufficient to resolve the complete spectrum for each Reynolds number.

    {Third, experimental data taken with a pinhole cap for the microphone suffer from Helmholtz resonance depending on the cavity between the microphone membrane and the cap. A common correction used is to fit an empirical model to calibration data and correct the resulting spectra with a transfer function. The model is not a perfect representation of the individual pinhole and perfect correction is nearly impossible. The result is an error in the spectra where the true cap effects are not perfectly accounted for.}

    {Fourth, considerations should be made when comparing experimental datasets due to the background noise rejection which is different in each facility. Background noise removal from coherent pressure fluctuations that are not a result of the boundary-layer physics can be done through various methods. The techniques of background noise removal can lead to discrepancies between experimental facilities in the low-frequency region.}
    
\begin{figure}
        \centering
        \includegraphics[width=0.86\textwidth]{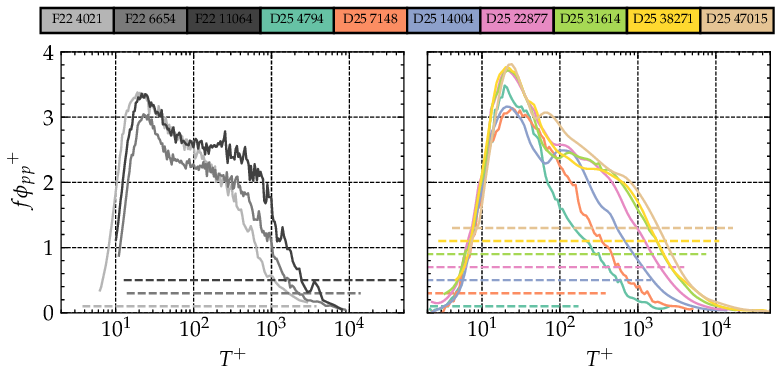} 
        \caption{Pre-multiplied spectra of wall-pressure fluctuations.  Wall pressure sensor frequency response given by horizontal bars, color corresponding to spectral data.  (a) Boundary layers for $\Rt=4021$  to 11064 \citep{fritsch_pressure_2020,fritsch_fluctuating_2022}.  (b) Pipe flows  for $\Rt=4794$  to 47\,015 \citep{dacome_scaling_2025}.   }
    \label{fig:filter} 
    \end{figure}

    Fifth,  the experimental data are limited by the frequency response of the wall pressure sensor.  
    Figure~\ref{fig:filter}a shows pre-multiplied spectra of wall-pressure fluctuations for the boundary layer data, with horizontal bars show range of $T^+$ corresponding to the frequency response of the pressure measurements (20Hz to 16 kHz).  Figure~\ref{fig:filter}b shows the corresponding limits for the pipe flow data (10 Hz to 40 kHz).  The boundary layer data at high $T^+$ (low frequencies) is well resolved at all three Reynolds numbers, but the data for $\Rt = 6654$ and 11\,064 appear to be significantly filtered at low $T^+$ (high frequencies).   The pipe flow data is affected somewhat in reverse, in that the low $T^+$ data is well resolved, but the high $T^+$ data is significantly filtered at the lowest three Reynolds numbers.  Despite the limitation on the frequency response, the effects of the filtering are relatively minor in terms of model presented here.

    Sixth, the $u_\tau^4$ dependence of the pre-multiplied spectra and the $u_\tau^2$ dependence on $T^+$ puts a great burden on measurements of $u_\tau$. The amplification in error affects both the height and $T^+$ position of the peak.
    

\bibliographystyle{jfm}
\bibliography{refs}
\end{document}